\documentclass[aps,pra,twocolumn,superscriptaddress,floatfix]{revtex4-2}%
\usepackage{graphicx}
\usepackage{color}
\usepackage{lipsum}
\usepackage{braket}
\usepackage{amsmath}
\usepackage{pifont}
\usepackage[colorlinks=true,citecolor=green,linkcolor=blue]{hyperref}
\usepackage{color}

\newcommand{\bt}{\textbf}

\begin{document}

\title{Telegraph flux noise induced beating Ramsey fringe in transmon qubits}
\author{Zhi-Hao~Wu}
\affiliation{College of Computer Science and Technology, National University of Defense Technology, Changsha 410073, China}
\author{Ling-Xiao~Lei}
\email{leilingxiao18@nudt.edu.cn}
\affiliation{College of Science, National University of Defense Technology, Changsha 410073, China}
\author{Xin-Fang~Zhang}
\email{xinfangzhang@quanta.org.cn}
\affiliation{College of Computer Science and Technology, National University of Defense Technology, Changsha 410073, China}
\author{Shi-Chuan~Xue}
\affiliation{College of Computer Science and Technology, National University of Defense Technology, Changsha 410073, China}
\author{Shun~Hu}
\affiliation{College of Computer Science and Technology, National University of Defense Technology, Changsha 410073, China}
\author{Cong~Li}
\affiliation{College of Computer Science and Technology, National University of Defense Technology, Changsha 410073, China}
\author{Xiang~Fu}
\affiliation{College of Computer Science and Technology, National University of Defense Technology, Changsha 410073, China}
\author{Ping-Xing~Chen}
\affiliation{College of Science, National University of Defense Technology, Changsha 410073, China}
\author{Kai~Lu}
\affiliation{College of Computer Science and Technology, National University of Defense Technology, Changsha 410073, China}
\author{Ming-Tang~Deng}
\email{mtdeng@nudt.edu.cn}
\affiliation{College of Computer Science and Technology, National University of Defense Technology, Changsha 410073, China}
\author{Jun-Jie~Wu}
\affiliation{College of Computer Science and Technology, National University of Defense Technology, Changsha 410073, China}

\begin{abstract}
Ramsey oscillations typically exhibit an exponential decay envelope due to environmental noise. However, recent experiments have observed nonmonotonic Ramsey fringes characterized by beating patterns, which deviate from the standard behavior. These beating patterns have primarily been attributed to charge-noise fluctuations. In this paper, we investigate the flux-noise origin of these nonmonotonic Ramsey fringes in frequency-tunable transmon qubits. We develop a random telegraph noise (RTN) model to simulate the impact of telegraph-like flux-noise sources on Ramsey oscillations. Our simulations demonstrate that strong flux-RTN sources can induce beating patterns in the Ramsey fringes, showing excellent agreement with experimental observations in transmon qubits influenced by electronic environment-induced flux-noise. Our findings provide valuable insights into the role of flux-noise in qubit decoherence and underscore the importance of considering flux-noise RTN when analyzing nonmonotonic Ramsey fringes.
\end{abstract}

\date{\today}

\maketitle

\section{Introduction}\label{introduction}

Ramsey oscillation is a fundamental phenomenon observed in quantum systems that allows for the investigation of qubit coherence and dynamics. The characteristics of Ramsey oscillations provide valuable insights into qubit coherence times and the interactions between qubits and their environments. For instance, the presence of strong environment noise can cause random fluctuations in qubit transition frequencies. These fluctuations result in qubit dephasing, which manifests as an exponential decay in the Ramsey oscillations. 

In addition to the monotonic decay envelope of the Ramsey oscillations, irregular Ramsey fringes have also been observed and documented in Refs.~\onlinecite{Riste2013, Stern2014, Peterer2015, Luthi2018, VonLupke2020, Ni2022, Martinez2023, Evert2024}. By analyzing deviations from ideal Ramsey oscillations, we can obtain insights into the nature and sources of noise in qubit systems, which is crucial for developing strategies to mitigate noise and enhance qubit performance. In these instances, Ramsey oscillations are often characterized by beating patterns. The beating patterns observed in Ramsey fringes are typically attributed to charge noises, such as charge drift or fluctuation, quasiparticle tunneling events through Josephson junctions, or parity switches in superconducting islands~\cite{Riste2013, Stern2014, Peterer2015, Luthi2018, Chang2022, Martinez2023}. Charge fluctuation can induce a doublet line structure in the qubit transition frequency, as observed in multiple studies~\cite{Bertet2005, Yoshihara2006, Schreier2008, Bal2015, Luthi2018}, thereby resulting in beating patterns in Ramsey oscillations.

For superconductor qubits such as transmon qubits with large ratios of $E_j/E_c$ (where $E_j$ represents the Josephson energy and $E_c$ denotes the charging energy), flux qubits and phase qubits, charge noise is significantly mitigated. In contrast, flux noise emerges as the predominant source of noise, particularly for qubits based on SQUID structures~\cite{Yoshihara2006, Koch2007, Bialczak2007, Kakuyanagi2007, Schreier2008, Bylander2011, Sank2012}. Flux noise originates from fluctuations in the magnetic flux threading through SQUID structures. External magnetic field variations and the motion of magnetic vortices are known to induce flux fluctuations. Notably, the most significant flux noise is low-frequency magnetic-flux noise, characterized by a $1/f$ power density spectrum, commonly referred to as the $1/f$ flux noise~\cite{Koch2007_noise, Bialczak2007, Zhou2012, Paladino2014, Falci2024}. Despite the proposal of several microscopic models for this $1/f$ flux noise, including surface/interface spin random reversal~\cite{Koch2007_noise}, diffusion~\cite{Lanting2014}, and clustering~\cite{De2014}, the exact origin of this phenomenon remains ambiguous.

%%%%%%%%%%%%%%%%%%%%%%%%%%%%%%%%%%%%%%%%%%%%%%%%
%%%%%%%%%%%% Figure-1 %%%%%%%%%%%%%%%%%%%
%%%%%%%%%%%%%%%%%%%%%%%%%%%%%%%%%%%%%%%%%%%%%%%
\begin{figure}
\centering
\includegraphics[width=8.5cm]{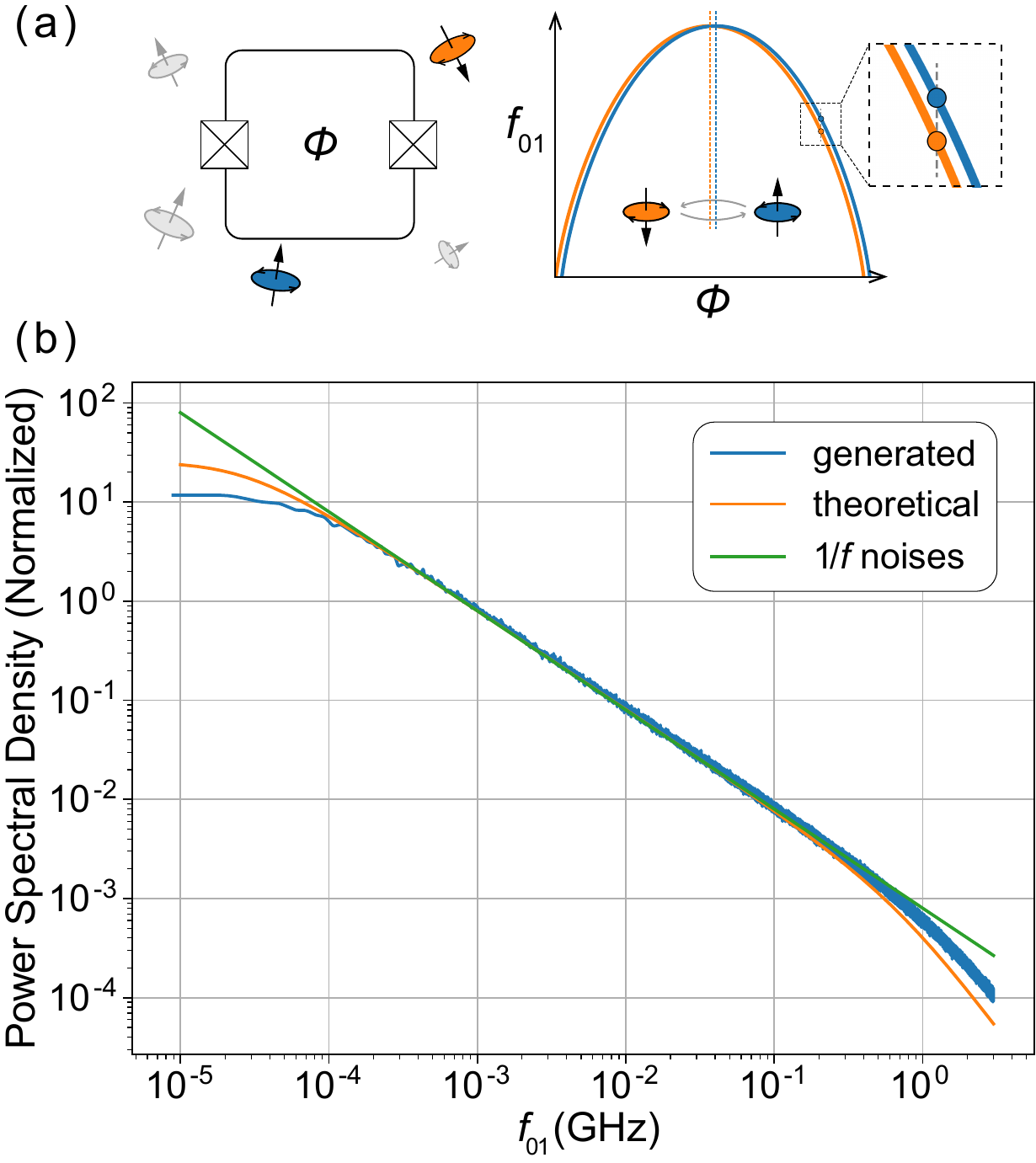}
\caption{\label{Fig1} \bt{Flux noise in frequency-tunable transmon qubits}. (\textbf{a}) Schematic illustration of a transmon qubit coupled to flux noise sources. A single strongly-coupled flux-RTN can induce a doublet-structure in the qubit transition frequency, leading to decoherence. (\textbf{b}) $1/f$ noise power spectrum density. The blue line represents the PSD obtained from the discrete Fourier transform of all generated $1/f$ noise sample functions. The orange line shows the theoretical PSD derived by summing the Lorentzian-type PSD contributions from each RTN source. The green line indicates the ideal PSD for $1/f$ noise. All three PSDs are normalized at $f_{01}=8\times 10^{-4}\ {\rm GHz}$.}
\end{figure}
%%%%%%%%%%%%%%%%%%%%%%%%%%%%%%%%%%%%%%%%%%%%%%%%%

Since flux noise represents one of the most significant sources of noise, it is pertinent to explore its potential role in inducing irregular Ramsey oscillations. Indeed, magnetic-field-dependent Ramsey beating has been sporadically observed. For example, Rower et al.~\cite{Rower2023} documented Ramsey beating patterns in the presence of a weak in-plane magnetic field. However, conclusive experimental evidence specifically attributing these phenomena to flux noise remains limited.

In this Letter, we investigate the flux-noise origin of nonmonotonic Ramsey fringes observed in frequency-tunable transmon qubits. We developed a decoherence model for transmon qubits and integrated it with a random telegraph noise (RTN) model to simulate Ramsey oscillations. Our simulations reveal that strong flux-RTN sources can cause beating patterns in Ramsey fringes, showing excellent agreement with experimental observations in transmon qubits influenced by flux-noise from the electronic environment. The remainder of this paper is organized as follows. In Sec.~\ref{secQubit}, we present the development of a decoherence model for transmon qubit devices. In Sec.~\ref{secRTN}, we introduce the RTN model. In Sec.~\ref{secRamsey}, we explore the integration of the qubit decoherence model and the RTN model into the numerical simulation of Ramsey oscillations, and compare these simulations with experimental results.

%%%%%%%%%%%%%%%%%%%%%%%%%%%%%%%%%%%%%%%%%%%%%%%%%%%%%
%%%%%% - Section - %%%%%%%
%%%%%%%%%%%%%%%%%%%%%%%%%%%%%%%%%%%%%%%%%%%%%%%%%%%%%%
\section{Decoherence Model of transmon qubit}\label{secQubit}

We first establish a decoherence model for a transmon qubit. Considering a single transmon qubit with a symmetric SQUID structure, the qubit transition frequency $\omega_{01}\equiv 2\pi f_{01}$ can be tuned by varying the external magnetic flux $\Phi_b$ applied to the SQUID. Under the two-level approximation, the Hamiltonian of the transmon qubit is given by
\begin{equation}
H_{\rm tra} = -\frac{\hbar \omega_{01}(\Phi_b)}{2} \sigma_z,
\end{equation}
where $\sigma_z$ represents the Pauli-z operator. For a transmon qubit, \( \omega_{01}(\Phi_b) \) can be approximately expressed as:
\begin{equation}\label{eq-phi-f01}
\omega_{01}(\Phi_b) = \frac{\sqrt{8E_c E_j |\cos(\pi \Phi_b/\Phi_0)|} - E_c}{\hbar},
\end{equation}
with $\Phi_0 \equiv \frac{h}{2e}$ denoting the magnetic flux quantum.

The decoherence effect in a quantum system is typically described by quantum master equations, which model the interaction between the system of interest and a large quantum environment, commonly referred to as a reservoir. The influence of the reservoir can be generalized by incorporating it into the quantum noise power spectral density (PSD) $S_Q(\omega)$ via the Bloch-Redfield equation \cite{bloch1957generalized, redfield1957theory, Breuer2002TheTO}:
\begin{equation}\label{eq-3}
\frac{d}{dt}\rho_{ab}(t) = -i\omega_{ba}\rho_{ab}(t) + \sum_{c,d} R_{abcd}\rho_{cd}(t),
\end{equation}
where $\rho_{ab}(t)$ denotes the element of the transmon qubit's density matrix, with the indices $a, b, c, d \in \left\{0, 1\right\}$ represent the matrix or tensor elements in the computation basis $\{|0\rangle, |1\rangle\}$, $\omega_{ba} \equiv (E_a - E_b)/\hbar$, and the 4-rank tensor $R_{abcd}$, known as the Bloch-Redfield tensor~\cite{Breuer2007,Vadimov_2021_PhysRevB}.

However, the Bloch-Redfield equation is derived under the Born-Markov approximation \cite{nielsen2010quantum,kolovsky2020quantum}, which is not appropriate for $1/f$ flux noise. This type of noise typically exhibits a long correlation time and its power spectrum diverges at $\omega = 0$~\cite{Paladino2014,Krantz2019}. A more suitable model for addressing low-frequency $1/f$ flux noise is to consider it as an adiabatic modulation of the qubit Hamiltonian. Specifically, $1/f$ flux noise and other low-frequency noises can be treated as classical stochastic processes [denoted as $\delta \Phi(t)$], which are incorporated into the external magnetic flux, leading to a modified Hamiltonian:
\begin{equation}\label{eq-4}
H_{\rm tra} = -\frac{\hbar \omega_{01} [\Phi_b + \delta \Phi(t)]}{2} \sigma_z.
\end{equation}
It is important to note that the stochastic process $\delta \Phi(t)$ is formally added to $\Phi_b$. In each dynamic process or individual experimental measurement, $\delta \Phi(t)$ should be replaced by a specific sample function $\delta \Phi(t, \varepsilon)$, where $\varepsilon$ is an element of the sample space $\mathcal{E}$.

Based on the time-dependent Hamiltonian presented in Eq.~\ref{eq-4}, we assume that the quantum system interacts with the reservoir exclusively via transverse coupling. Given that the PSD of quantum noise is approximately flat around the working bias $\Phi_b$ and $S_Q[-\omega_{01}(\Phi_b)] \ll S_Q[\omega_{01}(\Phi_b)]$, we can analytically solve Eq.~\ref{eq-3} to derive the evolution of the transmon qubit's density matrix in the Schr\"odinger picture as follows:
\begin{equation}\label{eq-5}
	\begin{aligned}
	\rho_{00}(t)&=\rho_{00}(0)\left\{1- \exp\left\{-\frac{S_Q[\omega_{01}(\Phi_b)]}{\hbar^2}t\right\}\right\},\\
	\rho_{01}(t)&=\rho_{01}(0) \exp\left\{{-\frac{S_Q[\omega_{01}(\Phi_b)]}{2\hbar^2}t+i\int_0^t\omega^\prime(\tau) d\tau}\right\},\\
	\rho_{10}(t)&=\rho_{10}(0) \exp\left\{{-\frac{S_Q[\omega_{01}(\Phi_b)]}{2\hbar^2}t-i\int_0^t\omega^\prime(\tau) d\tau}\right\},\\
	\rho_{11}(t)&=\rho_{11}(0) \exp\left\{-\frac{S_Q[\omega_{01}(\Phi_b)]}{\hbar^2}t\right\},
\end{aligned}
\end{equation} 
where $\omega^\prime(\tau) \equiv \omega_{01}[\Phi_b + \delta\Phi(\tau,\varepsilon)]$ represents the frequency of the transmon qubit during a specific dynamic process. Equation~\ref{eq-5} differentiates the decoherence effects induced by quantum noise from those caused by low-frequency noise. In a specific dynamic process, frequency fluctuations of the transmon qubit lead exclusively to phase accumulation in the off-diagonal elements of the density matrix.

A Ramsey experiment designed to extract decoherence information from off-diagonal elements is inherently a multi-shot experiment. Specifically, at each time point, the experiment is repeated multiple times to accurately determine the populations of $|0\rangle$ and $|1\rangle$. Each repetition corresponds to a particular sample function $\delta \Phi(t,\varepsilon)$. As the number of repetitions increases sufficiently, $\varepsilon$ effectively spans the entire sample space $\mathcal{E}$. Consequently, the off-diagonal elements in Eq.~\ref{eq-5} can be rewritten as
\begin{equation}\label{eq-6}
\begin{aligned}
\rho_{01}(t)&=\rho_{01}(0) e^{-\frac{S_Q[\omega_{01}(\Phi_b)]}{2\hbar^2}t}\cdot\left\langle e^{i\int_0^t\omega^\prime(\tau) d\tau}\right\rangle_{\varepsilon\in\mathcal{E}}\\
\rho_{10}(t)&=\rho_{10}(0) e^{-\frac{S_Q[\omega_{01}(\Phi_b)]}{2\hbar^2}t}\cdot\left\langle e^{-i\int_0^t\omega^\prime(\tau) d\tau}\right\rangle_{\varepsilon\in\mathcal{E}},
\end{aligned}
\end{equation}
where $\langle \cdot \rangle_{\varepsilon\in \mathcal{E}} = \frac{1}{|\mathcal{E}|}\sum_{\varepsilon\in \mathcal{E}} (\cdot)$. Compared with the diagonal elements in Eq.~\ref{eq-5}, Eq.~\ref{eq-6} successfully derives the well-known relation between the relaxation time $T_1$ and dephasing time $T_2$, given by
\begin{equation}
T_2=\frac{2T_1T_{2,\phi}}{2T_1+T_{2,\phi}},
\end{equation}
where $T_1 = \frac{\hbar^2}{S_Q[\omega_{01}(\Phi_b)]}$, and $T_{2,\phi}$ represents the decay time of $|\langle e^{-i\int_0^t \omega'(\tau) d\tau} \rangle_{\varepsilon \in \mathcal{E}}|$, assuming exponential decay.

Finally, we elucidate the connection between Eq.~\ref{eq-6} and the Ramsey experiment to facilitate subsequent discussions. In the context of the Ramsey experiment, let the detuning frequency be denoted as $\Delta\omega$. The Ramsey curve can then be expressed as 
\begin{equation}\label{eq-8}
P_1(t) = \frac{1}{2} \left[ 1 + \cos(\Delta \omega t) E(t) \right],
\end{equation}
where it is assumed that the qubit is initialized in the state $|0\rangle$. Here, $P_1(t)$ represents the population of the state $|1\rangle$ at time $t$, and the envelope function $E(t)$ is defined as the absolute value of the decay factor given by Eq.~\ref{eq-6}, specifically,
\begin{equation}\label{eq-9}
E(t) = e^{-\frac{S_Q[\omega_{01}(\Phi_b)]}{2\hbar^2}t} \cdot \left| \langle e^{i\int_0^t \omega'(\tau) d\tau} \rangle_{\varepsilon \in \mathcal{E}} \right|.
\end{equation}

Here, $E(t)$ represents the evolution of the modulus of the quantum state on the Bloch XY-plane. 

%%%%%%%%%%%%%%%%%%%%%%%%%%%%%%%%%%%%%%%%%%%%%%%%
%%%%%%%%%%%% Figure-2 %%%%%%%%%%%%%%%%%%%
%%%%%%%%%%%%%%%%%%%%%%%%%%%%%%%%%%%%%%%%%%%%%%%
\begin{figure*}
\centering
\includegraphics[width=16cm]{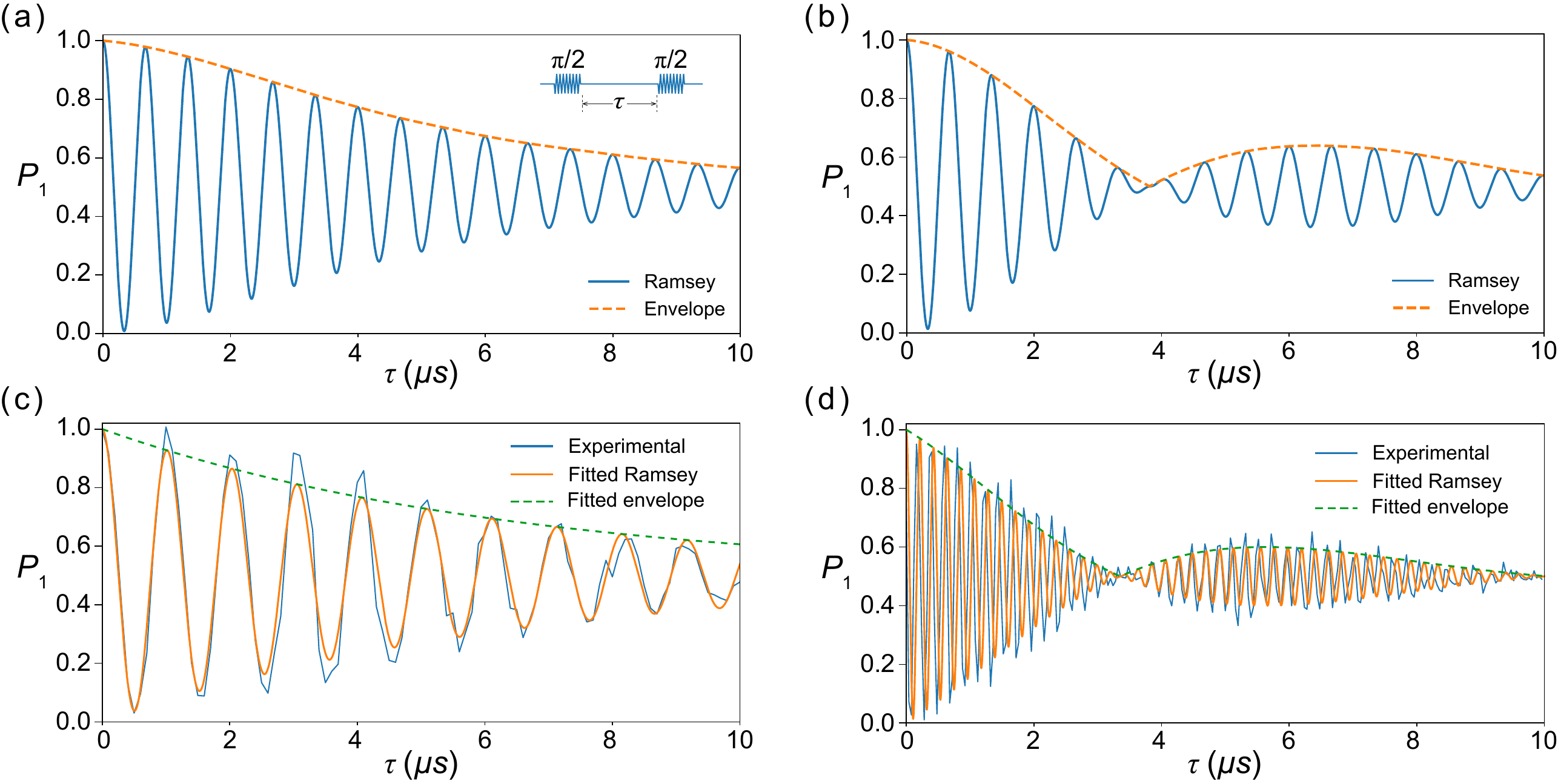}
\caption{\label{Fig2} \bt{Ramsey fringe of a transmon qubit}. 
(\textbf{a}) Ramsey oscillations of a transmon qubit simulated using an RTN flux noise model with a $1/f$ noise spectrum. The inset shows the pulse sequence of a typical Ramsey experiment. (\textbf{b}) Similar to panel (\textbf{a}), but incorporating an additional strong flux noise source characterized by a fixed switching frequency. The Ramsey fringes exhibit a beating pattern due to this additional noise component. (\bt{c}) Regular Ramsey oscillation fringe of a transmon qubit device (qubit-1). Blue line represents the experimental result, while yellow line depicts the Ramsey oscillation fitted by the function $(1+\cos(\Delta \omega \tau)\exp(-\Gamma \tau))/2$, and green line highlights the envelope $\exp(-\Gamma t)$.  (\textbf{d}) Similar to panel (\textbf{c}) but with a beating envelope. The fitting function is modified to $(1+\cos(\Delta \omega \tau)\exp(-\Gamma \tau)|\cos(\delta \omega \tau/2)|)/2$ according to Eq.~\ref{eq-15}. The frequency detuning $\Delta\omega$ in panel (\bt{d}) is set to a larger value compared to that in panel (\bt{c}) in order to more clearly display the beating pattern.}
\end{figure*}
%%%%%%%%%%%%%%%%%%%%%%%%%%%%%%%%%%%%%%%%%%%%%%%%%

%%%%%%%%%%%%%%%%%%%%%%%%%%%%%%%%%%%%%%%%%%%%%%%%%%%%%
%%%%%% - Section - %%%%%%%
%%%%%%%%%%%%%%%%%%%%%%%%%%%%%%%%%%%%%%%%%%%%%%%%%%%%%%
\section{RTN model}\label{secRTN}

To generate the noise PSD for Eq.~\ref{eq-9} as discussed in the previous section, it is necessary to develop a noise model capable of producing $1/f$ noise. $1/f$ noise can be understood from various perspectives. One of the most widely accepted approaches is to interpret $1/f$ noise as resulting from the superposition of a large number of bistable fluctuators \cite{Paladino2014, Kogan1984}. The noise contribution from each fluctuator is characterized by random telegraph noise \cite{Machlup1954, Kirton1989, Kogan1996}.

Regarding $1/f$ flux noise, these fluctuators can be attributed to surface or interface spin random reversals or other physical phenomena discussed in Section \ref{introduction}. As illustrated in left panel of Fig.~\ref{Fig1}(a), these fluctuators are distributed throughout the quantum chip hosting transmon qubits. The two stable states of a single fluctuator will induce either a positive or negative flux in the SQUID, both with equal magnitude. Given that electron spins are relatively insensitive to the surrounding environment, it is reasonable to assume that the transition rates between the two stable states of a fluctuator are identical (i.e., $\lambda_{\uparrow \rightarrow \downarrow} = \lambda_{\downarrow \rightarrow \uparrow} \equiv \lambda$). Mathematically, the flux contribution from an individual fluctuator can be expressed as an RTN, which reads
\begin{equation}\label{eq-10}
	{\rm RTN}(t)= s_0b(-1)^{N(t)},
\end{equation} 
where $s_0$ is a uniform random variable with $P(s_0 = -1) = P(s_0 = +1) = 1/2$, determining the initial state of the fluctuator at $t = 0$, $b$ represents the amplitude of the RTN, and $N(t)$ denotes a Poisson random process with a switching frequency $\lambda$.

The RTN noise described in Eq.~\ref{eq-10} is a wide-sense stationary stochastic process with a mean value of \( E[\text{RTN}(t)] = 0 \) and an autocorrelation function given by \( R_{\text{RTN}}(|s-t|) = E[\text{RTN}(t)\text{RTN}(s)] = e^{-\lambda|t-s|} \) for \( t, s \geq 0 \). By applying the Wiener-Khinchin theorem~\cite{Beichlt2002}, the power spectral density of the RTN noise can be expressed as a Lorentzian function:
\begin{equation}\label{eq-Lorentzian}
	S_{\rm RTN}(\omega)=\frac{2\lambda}{\omega^2+\lambda^2}.
\end{equation}  

Assuming there are $N$ fluctuators on the quantum chip, each with an identical magnitude $b$ and negligible interaction between them, the total noise and its power spectral density can be expressed as
\begin{equation}\label{eq-12}
\begin{aligned}
	{\rm Flicker(t)}=\sum_{i=0}^N {\rm RTN}_i(t)\quad 
	S_{\rm Flicker}(\omega)=\sum_{i=0}^N S_{{\rm RTN}_i}(\omega).
\end{aligned}	
\end{equation}
If $N$ is sufficiently large and the switching frequency of all fluctuators $\left\{\lambda_i|i=1,\cdots,N\right\}$ follow the distribution $P(\lambda) \propto 1/\lambda$~\cite{Paladino2014}, the summation in Eq.~\ref{eq-12} can be approximated by an integral:
\begin{equation}\label{eq-13}
	\begin{aligned}
	S_{\rm Flicker}(\omega)&= \int_{\lambda_{\rm min}}^{\lambda_{\rm max}}P(\lambda) S_{{\rm RTN}_i}(\omega) d\lambda\\
	&=\frac{2b^2P_0\left(\arctan(\lambda/\omega)|_{\lambda_{\rm min}}^{\lambda_{\rm max}}\right)}{\omega},
	\end{aligned}
\end{equation} 
where $P_0 = N / \ln(\lambda_{\rm max} / \lambda_{\rm min})$ is the normalization factor of $P(\lambda)$, $\lambda_{\rm min}$ ($\lambda_{\rm max}$) is the minimal (maximal) cutoff switching frequencies of those fluctuators. The frequency range we studied typically satisfies the condition $\lambda_{\rm min} \ll \omega \ll \lambda_{\rm max}$, implying that Eq.~\ref{eq-13} can be approximated as 
\begin{equation}\label{eq-1/f}
	S_{\rm Flicker}(\omega)\approx \frac{ \pi b^2 P_0}{\omega}\propto \frac{1}{\omega},
\end{equation}
which conforms to the definition of $1/f$ noise. 

To simulate strong coupled flux noise,  an individual RTN with a substantially  higher amplitude can be superimposed on the $1/f$ noise spectrum. This additional noise component directly influences the $f_{01}$ spectrum, leading to a doublet structure as illustrated in the right panel of Fig.~\ref{Fig1}(a). Assuming the spectral splitting between the two stable states at the working bias $\Phi_b$ is $\delta \omega$, the modified Ramsey curve approximates to
\begin{equation}\label{eq-15}
	\begin{aligned}
	P_1(t) \approx \frac12 \left[ 1+\cos(\Delta \omega t)|\cos(\delta \omega t/2 )|E(t)\right],
	\end{aligned}
\end{equation}  
Eq.~\ref{eq-15} implies a new Ramsey envelope $E^\prime(t) = |\cos(\delta \omega/2 t)| E(t)$, which directly indicates the Ramsey beating phenomenon.

Eq.~\ref{eq-15} could be derived strictly through the analytic discussion of decay factor $\langle e^{-i\int_0^t\omega^\prime(\tau) d\tau}\rangle_{\varepsilon\in\mathcal{E}}$ in Appendix.~\ref{AppendixAnalysis}. Specifically, Eq.~\ref{eq-C5} illustrates that the new Ramsey envelope $E^\prime(t)$ reads
\begin{equation}
	E^\prime(t)= \left|\left\langle \exp\left(i\int_0^t\frac{d \omega_{01}}{d\Phi_b}\big|_{\Phi_b}{\rm RTN}(\tau,\varepsilon_j) d\tau\right) \right\rangle_{\varepsilon\in \mathcal{E}^\prime}\right|E(t).
\end{equation}
Assuming the switching frequency $\lambda$ is sufficiently low, the probability that the fluctuator does not switch during a single experiment ,i.e. $P(n=0)$, would be close to one. For example, if a single experiment lasts for 50 $\rm \mu s$ and $\lambda=50\ {\rm Hz}$, then $P(n=0)= 0.9975$. Regarding Eq.~\ref{eq-C9}, the preceding discussion indicates that the expansion can be truncated to the $n=0$ term, implying
\begin{equation}\label{eq-17}
	E^\prime(t)=e^{-\lambda t}\left|\cos\left(\frac{d\omega_{01}}{d\Phi_b}\bigg|_{\Phi_b}\cdot b\cdot t\right)\right|E(t).
\end{equation}
Since \( b \) remains significantly smaller than \( \Phi_0 \), we have $d\omega_{01}/d\Phi_b|_{\Phi_b}\cdot b\approx \delta \omega/2$. Additionally, the extra decay factor \( e^{-\lambda t} \) is negligible given that \( \lambda \) is sufficiently small. Consequently, we have derived Eq.~\ref{eq-15}.

In addition to the case caused by a single prominent RTN, we can also predict other nonmonotonic Ramsey oscillations resulting from a few prominent RTNs. Specifically, if there are \( N_s \) additional RTNs with amplitudes \( \{b_j\} \), the Ramsey envelope \( E^{\prime\prime}(t) \) is given by
\begin{equation}
	E^{\prime\prime}(t)=\prod_{j=1}^{N_s}\left|\cos\left(\frac{d\omega_{01}}{d\Phi_b}\bigg|_{\Phi_b}\cdot b_j \cdot t\right)\right|E(t).
\end{equation} 
%%%%%%%%%%%%%%%%%%%%%%%%%%%%%%%%%%%%%%%%%%%%%%%%%%%%%
%%%%%% - Section - %%%%%%%
%%%%%%%%%%%%%%%%%%%%%%%%%%%%%%%%%%%%%%%%%%%%%%%%%%%%%%

\section{Ramsey oscillation simulation and experiment}\label{secRamsey}
Based on the decoherence and RTN models described in Sec.~\ref{secQubit} and Sec.~\ref{secRTN}, we can simulate the Ramsey oscillation by numerically evaluating the term $\langle e^{-i\int_0^t\omega^\prime(\tau) d\tau}\rangle_{\varepsilon\in\mathcal{E}}$ in Eq.~\ref{eq-9}, assuming that the PSD of quantum noise $S_Q(\omega)$ remains constant across the entire tunable frequency range of the transmon qubit. The details of this numerical simulation is described in Appendix~\ref{AppendixRamsey}. As illustrated in Fig.~\ref{Fig1}(b), the PSD generated from the RTN model with the sampling process  aligns well with both the theoretical summation of Lorentzian functions (Eq.~\ref{eq-Lorentzian} and Eq.~\ref{eq-12}) and the ideal $1/f$ spectrum (Eq.~\ref{eq-1/f}).

By combining the RTN-generated PSD with the decoherence model of the transmon qubit, we can numerically simulate Ramsey oscillations. As shown in Fig.~\ref{Fig2}(a), the simulated Ramsey fringe envelope exhibits the expected exponential decay.

To investigate the flux-noise-induced nonmonotonic Ramsey fringe, we introduce one or a few RTN noise sources with significantly higher amplitudes into the stochastic $1/f$ process $\delta\Phi(t)$. As shown in Fig.~\ref{Fig2}(b), when a strongly-coupled flux-RTN is present, the Ramsey oscillation pattern exhibits a modified behavior characterized by a beating envelope. Specifically, for the strongly-coupled flux-RTN, we set the switching frequency to \(\lambda = 50\ \text{Hz}\) and the amplitude to \(b = 4.2 \times 10^{-5}\ \Phi_0\). For the simulated transmon qubit, we set the bias flux to \(\Phi_b = -0.06051\ \Phi_0\).

We have experimentally observed Ramsey fringes with both monotonic decaying and beating envelopes in a superconducting quantum circuit chip featuring frequency-tunable transmon qubits (Appendix~\ref{AppendixSetup}). By performing the standard Ramsey experiment on one of the transmon qubits, we observed a typical exponentially decaying Ramsey fringe pattern [Fig.~\ref{Fig2}(c)], which indicates that the qubit decoherence is primarily governed by the aforementioned stochastic process. However, when one of the measurement instruments' location was switched to an alternative position (the ``noisy spot") (Fig.~\ref{FigS3}), a beating pattern emerged superimposed on the Ramsey fringe [Fig.~\ref{Fig2}(d)]. We fitted both the exponential decay and the beating patterns using analytical functions derived from the RTN model, achieving a highly consistent match with the experimental data.

%%%%%%%%%%%%%%%%%%%%%%%%%%%%%%%%%%%%%%%%%%%%%%%%
%%%%%%%%%%%% Figure-3 %%%%%%%%%%%%%%%%%%%
%%%%%%%%%%%%%%%%%%%%%%%%%%%%%%%%%%%%%%%%%%%%%%%
\begin{figure}
\centering
\includegraphics[width=8.5cm]{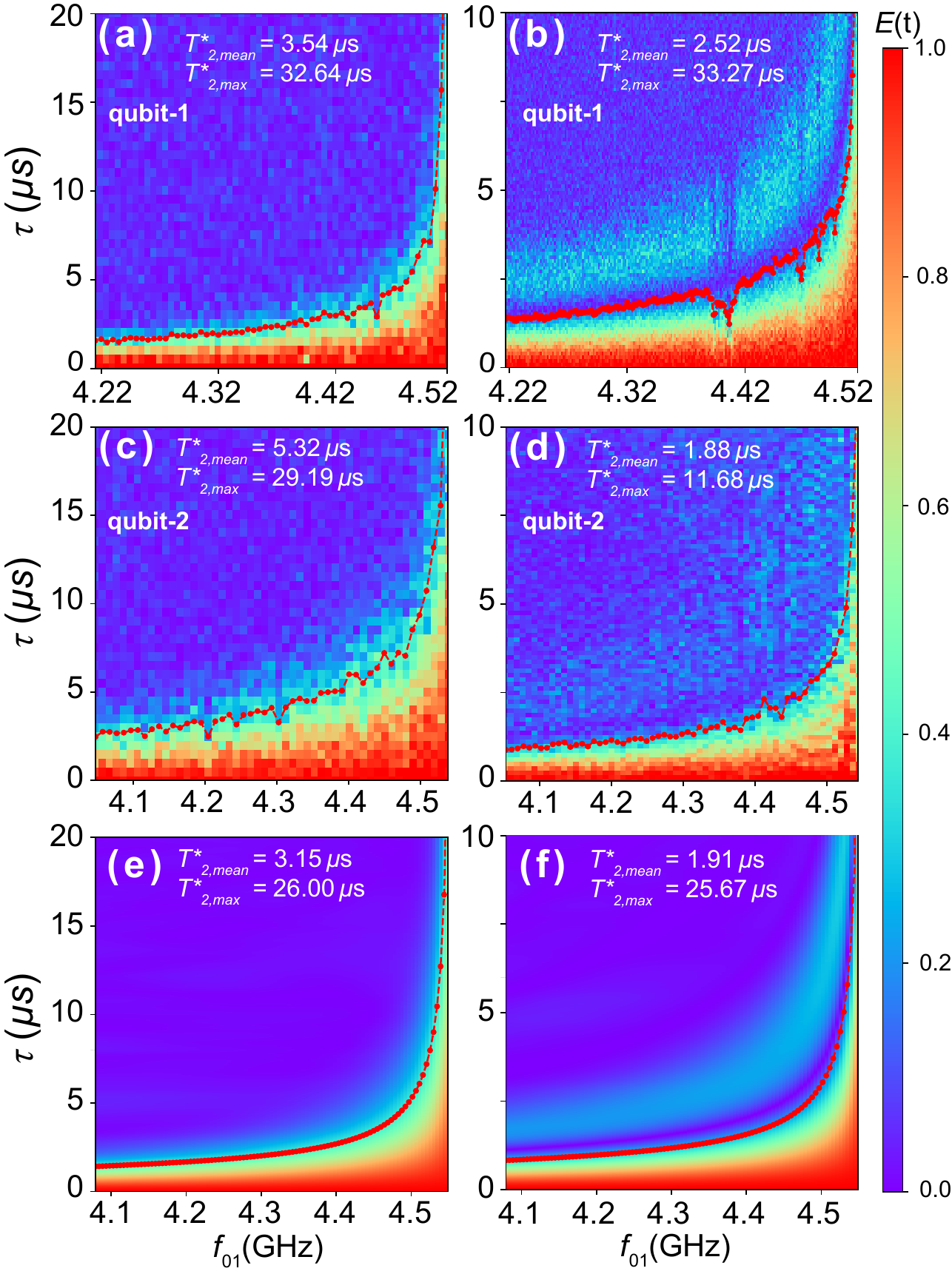}
\caption{\label{Fig3} \bt{Extracted qubit state vector XY-modulus (i.e., $E(t)$ or  $E^\prime(t)$) from the Ramsey experiment as a function of qubit frequency $f_{01}$ and the time interval $\tau$.} (\bt{a}) and (\bt{c}) show experimental results for qubit-1 and qubit-2. For both qubits, the Ramsey fringe envelope exhibits exponential decay at all frequencies. Red dots represent the decoherence time $T_2^*$ obtained by fitting the Ramsey envelopes, with the mean and maximum values of $T_2^*$ labeled. $T_2^*$ decreases significantly when the qubit is far from the optimal point. (\bt{b}) and (\bt{d}) are similar to panels (\bt{a}) and (\bt{c}), but they exhibit a beating phenomenon across the entire tunable frequency range. (\bt{e}) shows the numerically simulated $E(t)$ spectrum when only $1/f$ noise is applied to the transmon qubit. (\bt{f}) is similar to panel (\bt{e}) but includes an additional strong RTN. The simulations are in excellent agreement with the experimental results.}
\end{figure}
%%%%%%%%%%%%%%%%%%%%%%%%%%%%%%%%%%%%%%%%%%%%%%%%%

To further investigate the beating Ramsey envelope, we measured the frequency-dependent Ramsey oscillations for two transmon qubits, specifically qubit-1 and qubit-2 (Fig.~\ref{FigS2} illustrates the behavior of the other three qubits, which exhibit similar characteristics). As illustrated in Figs.~\ref{Fig3} (a)-(d), the $T_2^*$-time of the qubits exhibits a strong dependence on frequency, which is influenced by the derivative $df_{01}/d\Phi_b$ (see Eq.~\ref{eq-phi-f01}). The beating pattern emerges and only emerges when the measurement instrument (shown in Fig.~\ref{FigS3}) is positioned at the ``noisy spot". This beating pattern spans the entire frequency range of the qubits. Additionally, the frequency of the beating pattern depends on $f_{01}$, with faster beating observed when $df_{01}/d\Phi_b$ is larger (i.e., when $f_{01}$ is smaller). Figures~\ref{Fig3} (e) and (f) present numerical simulations based on the flux noise RTN model, without and with additional strong flux-RTN, respectively. The excellent agreement between simulation and the measurement rules out the coherently coupled two-level-system (with energy level avoided-crossing~\cite{Martinis2005}) or the charge-noise~\cite{Schreier2008, Riste2013, Bal2015} as the mechanism of the beating Ramsey.

From the above investigation, we can attribute the observed beating pattern to one or a few telegraph-like flux-noise sources that are strongly coupled to the qubits. A plausible mechanism for this flux noise is electromagnetic radiation emitted at a specific frequency by nearby instrumentation. This radiation can be picked up by the plugboard and transmitted through coaxial cables to the chip, thereby affecting all qubits on the chip. The impact of this radiation would likely be significantly mitigated if the instrument were relocated away from the plugboard.

%%%%%%%%%%%%%%%%%%%%%%%%%%%%%%%%%%%%%%%%%%%%%%%%
%%%%%%%%%%%% Figure-4 %%%%%%%%%%%%%%%%%%%
%%%%%%%%%%%%%%%%%%%%%%%%%%%%%%%%%%%%%%%%%%%%%%%
\begin{figure}
\centering
\includegraphics[width=8.5cm]{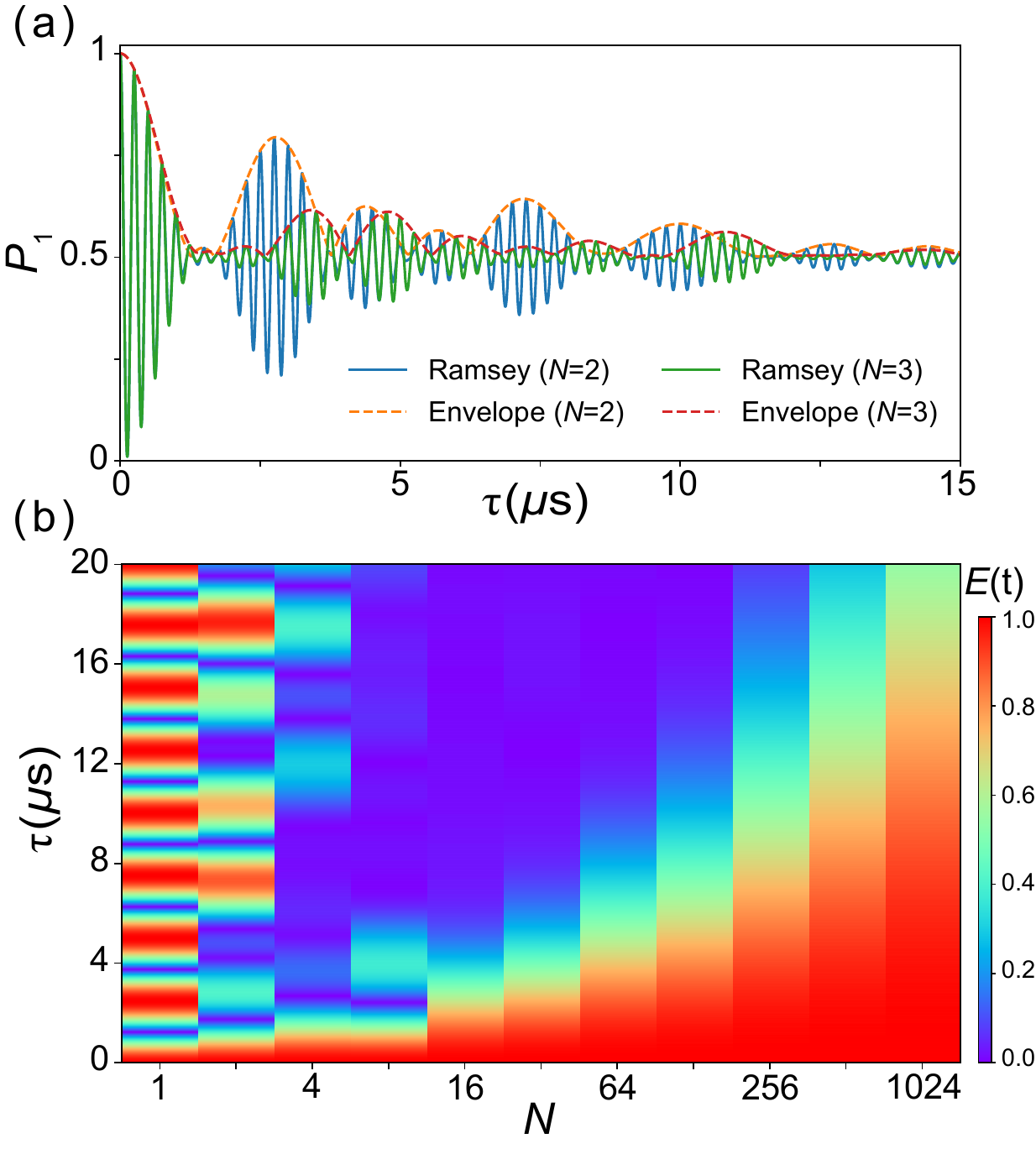}
\caption{\label{Fig4} \bt{Ramsey fringes induced by multiple strong flux-RTN sources.} (\textbf{a}) Ramsey fringes in the presence of $1/f$ noise and $N$ flux-RTN sources, where $N = 2, 3$. \textbf{(b)} Simulated $E(t)$ for Ramsey process as a function of $N$. From left to right, the amplitude of the additional flux-RTN is distributed among $N$ individual sources. For each $N$, the amplitudes of individual flux-RTN sources $b_i (i=1,2,\ldots,N)$ are randomly generated while their total sum remains constant, i.e., $\sum\limits_{i=1}^N{b_i}=b_0$. The spectra demonstrate how the beating pattern evolves into a typical exponential decay pattern as the influence of the additional strong flux-RTN is averaged out. ($\lambda=50\ {\rm Hz}$, $b_0=8\times 10^{-5}\ \Phi_0$,  $\Phi_b=0.0966\ \Phi_0$)}
\end{figure}
%%%%%%%%%%%%%%%%%%%%%%%%%%%%%%%%%%%%%%%%%%%%%%%%%

If multiple strong RTNs with varying amplitudes are present, the Ramsey fringe envelope becomes increasingly irregular. As illustrated in Fig.~\ref{Fig4}(a), Ramsey fringes with $N=2$ and $N=3$ strong RTNs exhibit complex beating patterns, and the damping of Ramsey oscillations increases as $N$ grows. In fact, when noise sources transition from a ``strong-but-few" scenario to a ``weak-but-many" scenario, the Ramsey fringe pattern evolves from a beating pattern to an exponentially decaying one [Fig.~\ref{Fig4}(b)].

\section{Conclusion}

In this paper, we have observed beating Ramsey fringes in frequency-tunable transmon qubits. The beating pattern is found to be dependent on the qubit frequency and influenced by the positioning of an electronic instrument. We attribute the non-monotonic Ramsey pattern to electromagnetic radiation noise originating from the instrument. To model this phenomenon, we developed a flux-RTN model and simulated the beating Ramsey fringes, achieving excellent agreement with the experimental data. Our results demonstrate that telegraph-like flux noise can induce a beating Ramsey pattern similar to charge noise. Although the flux noise in this experiment originates from a macroscopic environment, our findings can be extended to microscopic flux noise, providing insights into analyzing qubit decoherence channels~\cite{Braumuller2020, Zhang2025} through Ramsey experiments.

\vspace{5pt}
\bt{Acknowledgments.} We would like to express our gratitude to Yang Yang, Xiao-Feng Yi, Yi-Chuan Zeng, and Wei-Chen Wang for their technical assistance with qubit measurements. We also thank Shao-Jun Guo and Wen-Hao Chu for insightful discussions. This work was supported by the National Key Research and Development Program of China (Grant No. 2024YFB4504001).

\vspace{5pt}
\textbf{Contribution.} Z.-H.W., X.-F.Z., S.H., X.F., S.-C.X., and M.-T.D. conducted the qubit experiments; Z.-H.W., L.-X.L., S.H., C.L., and M.-T.D. analyzed the experimental data; L.-X.L. developed the RTN model with assistance from S.H.; L.-X.L. developed the qubit decoherence model and simulated qubit Ramsey oscillations; P.-X.C., K.L., M.-T.D., and J.-J.W. supervised both the experiments and simulations; M.-T.D. initiated and led the project. All authors contributed to the result discussion and manuscript writing.

\appendix 
\renewcommand{\thefigure}{S\arabic{figure}}
\setcounter{figure}{0}

\section{Experiment setup}
\label{AppendixSetup}

The transmon qubits from qubit-1 to qubit-5 are from a superconducting circuit chip featuring a structure analogous to those reported in Refs.~\onlinecite{Arute2019, Wu2021}. On the chip, the frequency-tunable transmon qubits are interconnected by a coupling-tunable coupler qubit between adjacent transmon qubits, as described in Yan et al.~\cite{Yan2018}. The qubit readout is implemented by a dispersive measurement scheme with Josephson parameter amplifiers and high electron mobility transistor (HEMT) amplifiers.

%%%%%%%%%%%%%%%%%%%%%%%%%%%%%%%%%%%%%%%%%%%%%%%%
%%%%%%%%%%%% Figure-A3 %%%%%%%%%%%%%%%%%%%
%%%%%%%%%%%%%%%%%%%%%%%%%%%%%%%%%%%%%%%%%%%%%%%
\begin{figure}
\centering
\includegraphics[width=6cm]{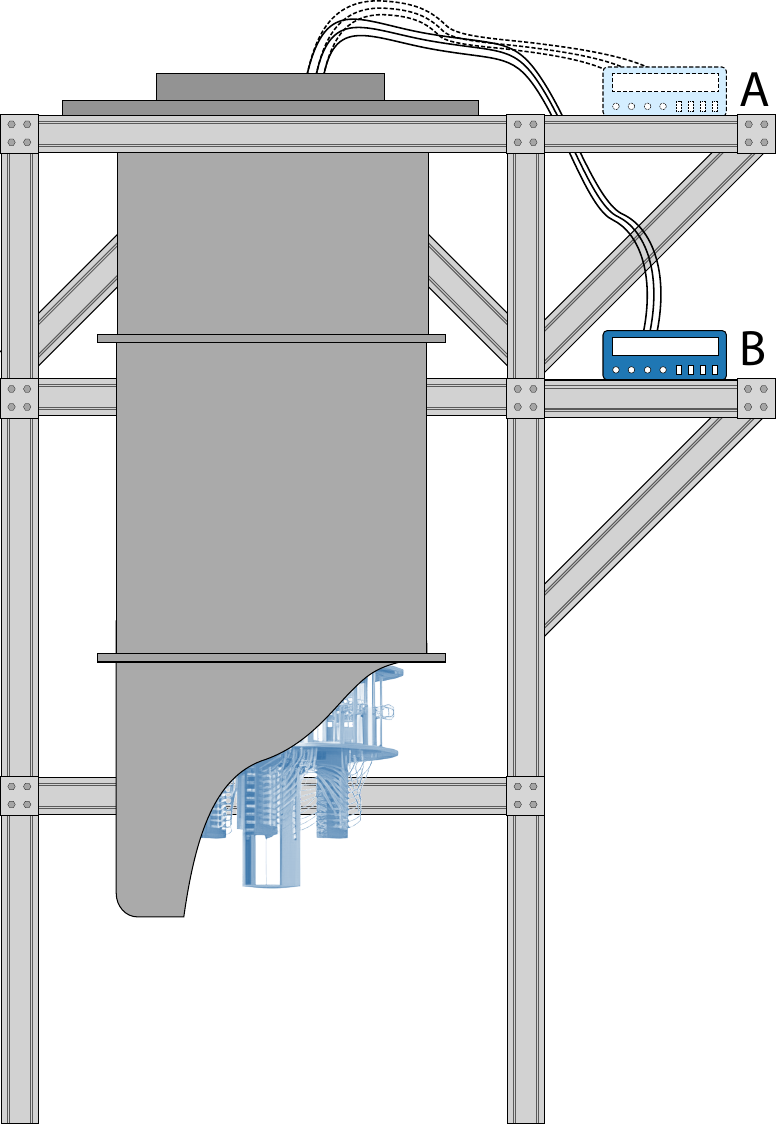}
\caption{\label{FigS3} \bt{Sketch of the measurement setup}. The superconducting qubit chip is measured in a dilution refrigerator with a base temperature of approximately 20 mK. Each readout channel consists of a HEMT amplifier powered by a DC-current source. The beating Ramsey interference pattern appears when the current source is positioned at location A, and disappears when the current source is placed at location B.}
\end{figure}
%%%%%%%%%%%%%%%%%%%%%%%%%%%%%%%%%%%%%%%%%%%%%%%

\begin{figure}
\centering
\includegraphics[width=8cm]{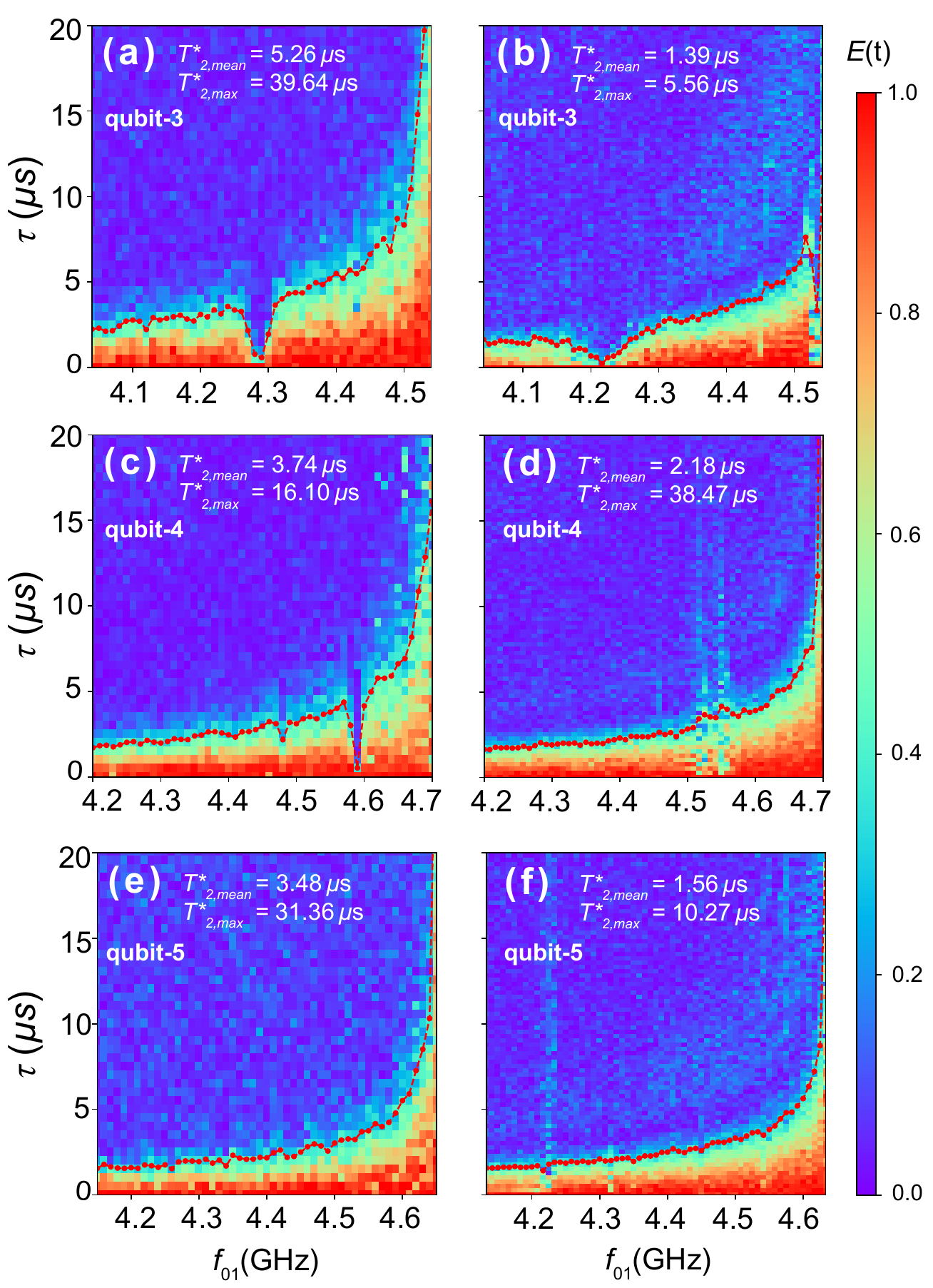}
\caption{\label{FigS2} \bt{Experiment Ramsey oscillation measurements from qubit-3, qubit-4 and qubit-5.} Similar to qubit-1 and qubit-2, all of the three qubits exhibit monotonic decaying patterns when the DC-current instrument is placed at location B [panels (\bt{a}), (\bt{c}) and (\bt{e})]. In contrast, beating behaviors are observed for all three qubits when the instrument is placed at location A [panels (\bt{b}), (\bt{d}) and (\bt{f})].}
\end{figure}

The flux noise investigated experimentally in this work originates from DC-current source instruments used to power HEMT amplifiers. As illustrated in Fig.~\ref{FigS3}, the flux noise causing beating in the Ramsey interference pattern is prominent when the DC-current source is positioned at location A. Conversely, when the current source is placed at location B, the beating patterns in the Ramsey interference are eliminated. A potential mechanism for the flux noise is the electromagnetic radiation emitted at a specific frequency by the DC current-source instrument. This radiation can be picked up by the coaxial cable and transmitted through the plugboard to the chip. The absorption of this radiation would be significantly reduced if the instrument was relocated away from the top of the dilution refrigerator (to location B). In addition to qubit-1 and qubit-2 discussed in the main text, many other qubits on the chip exhibit similar behaviors, which are contingent upon the instrument locations (Fig.~\ref{FigS2}).

%%%%%%%%%%%%%%%%%%%%%%%%%%%%%%%%%%%%%%%%%%%%%%%%
%%%%%%%%%%%% Figure-A2 %%%%%%%%%%%%%%%%%%%
%%%%%%%%%%%%%%%%%%%%%%%%%%%%%%%%%%%%%%%%%%%%%%%
%%%%%%%%%%%%%%%%%%%%%%%%%%%%%%%%%%%%%%%%%%%%%%%

\section{Details of Ramsey oscillation simulations}
\label{AppendixRamsey}

In this section, we present the detailed numerical simulations of Ramsey oscillations using both the qubit coherence model and the flux-RTN model.

The numerical simulations is implemented as follows.
\begin{itemize}
\item \textbf{Step-I}: RTN generation. For each RTN, generate 3000 sample functions. The switching frequencies of these RTN follow the distribution $P(\lambda) \propto 1/\lambda$.

\item \textbf{Step-II}: Flicker noise generation. Add the sample functions of RTNs to generate the $1/f$ noise ${\rm Flicker}(t)$ function as described in Eq.~\ref{eq-12}.

\item \textbf{Step-III}: Decay factor calculation. Employ numerical integration to evaluate the decay factor $\langle e^{i\int_0^t \omega'(\tau) d\tau }\rangle_{\varepsilon\in \mathcal{E}}$ as presented in Eq.~\ref{eq-9}. In this computation, the noise term $\delta \Phi(t)$ from Eq.~\ref{eq-4} is substituted with ${\rm Flicker}(t)$ obtained from the previous step, and the discrete sampling interval $\delta t$ is specified as $0.2\ {\rm ns}$.

\item \textbf{Step-IV}: Combine the PSD of quantum noise $S_{Q}(\omega_b)$ and the detuning frequency $\Delta \omega$ to derive the Ramsey oscillation curve presented in Eq.~\ref{eq-8}.
\end{itemize}

%$S_{Q}(\omega_b)= \hbar^2/(13\ {\rm \mu s})$
%$\Delta \omega= 1.5\ {\rm MHz}$
The methods employed in most steps are well-established and can be implemented directly. However, the generation of RTN sample functions is more intricate and less straightforward. Therefore, we will elaborate on the process of generating RTN sample functions in detail in the following discussion.

Assuming the total simulation time is \( T \), the sample function of an RTN is determined by the number of switching events \( N(T) \) between two stable states of the fluctuator during the interval \([0, T]\), the initial electron spin direction \( s_0 \) at \( t = 0 \), and the specific time points at which these \( N(T) \) switches occur (denoted as \( \{\mathcal{T}_i | 1 \leq i \leq N(T)\} \) with \( \mathcal{T}_i \leq \mathcal{T}_j \) for any \( i \leq j \)). These quantities are all random variables, where \( N(T) \) follows a Poisson distribution with arrival rate \( \lambda \), \( s_0 \) is uniformly distributed, and \( \{\mathcal{T}_i | 1 \leq i \leq N(T)\} \) are independent and identically distributed random variables, each uniformly distributed over the interval \([0, T]\)~\cite{Beichlt2002}. 

Based on the above discussion, we can create RTN sample functions through the following steps:
\begin{itemize}
\item \textbf{Step-I}: Utilize random numbers with the Poisson and uniform distributions to generate $N(T)$ and $s_0$, respectively.
  
\item \textbf{Step-II}: Generate $n$ random numbers with a uniform distribution over the interval $[0, T]$, where $n$ corresponds to the specific realization of the random variable $N(T)$ obtained in \bt{Step-I}.
  
\item \textbf{Step-III}: Calculate the sample function (i.e. ${\rm RTN}(t,s_0^\prime,n,\left\{t_i\right\})$) according to the definition of RTN given in Eq.~\ref{eq-10}. Here, $s_0^\prime$, $n$, and $\left\{t_i\right\}$ represent specific realizations of the random variables $s_0$, $N(T)$, and $\left\{\mathcal{T}_i\right\}$, respectively.
\end{itemize}

%Specifically, we set the Josephson energy to \( E_j/\hbar = 13.5\ \text{GHz} \) and the charge energy to \( E_c/\hbar = 0.1\ \text{GHz} \). As a result, the frequency of the transmon qubit in our simulation can be tuned within the range of \( [0, 4.5475]\ \text{GHz} \). Additionally, we set the amplitudes \( \{b_i\} \) of all 3000 RTNs to \( 5/3 \times 10^{-6}\ \Phi_0 \). 

\section{Analytic discussions on Ramsey oscillation} \label{AppendixAnalysis}

Although numerical simulations of Ramsey oscillation have provided an efficient method for predicting the decoherence of transmon qubits and nonmonotonic Ramsey fringes, further analytic discussions are necessary to explore the detailed characteristics of nonmonotonic Ramsey fringes. In this section, we present the analytic solution to the decay factor in Eq.~\ref{eq-9}. 

Based on the definition of low-frequency noise $\delta \Phi_0(t)$, this noise should be a small quantity relative to the flux quantum $\Phi_0$. Therefore, Consequently, the time-dependent qubit frequency $\omega^\prime(t)$ can be approximated as a linear function, which reads
\begin{equation}\label{eq-C1}
	\omega^\prime (t) = \omega_{01}(\Phi_b)+\frac{d \omega_{01}}{d \Phi_b}\bigg|_{\Phi_b} \delta \Phi(t,\varepsilon), 
\end{equation} 

Assuming the low-frequency noise $\delta \Phi(t)$ comprises $N$ RTNs of identical amplitude $b$ from the $1/f$ noise source ${\rm Flicker}(t)$ and $N_s$ prominent RTNs ${\rm RTN}_i (t)$ ($i=1,\cdots,N_s$) with distinct amplitudes $\left\{b_i\right\}$, which means 
\begin{equation}\label{eq-C2}
\begin{aligned}
	\delta\Phi(t) = {\rm Flicker}(t) +\sum_{i=1}^{N_S} {\rm RTN}_i(t).
\end{aligned}
\end{equation}
Moreover, although the amplitudes vary, the $N_s$ RTNs are not fundamentally different. Consequently, we can hypothesize that the sample space of all RTNs is $\mathcal{E}^\prime$. Therefore, the total sample space can be expressed as
\begin{equation}\label{eq-C3}
	\begin{aligned}
	\mathcal{E}&=\mathcal{E}_{1/f}\times \underbrace{\mathcal{E}^\prime\times \cdots \times \mathcal{E}^\prime}_{N_s}\\
	&=\underbrace{\mathcal{E}^\prime\times \cdots \times \mathcal{E}^\prime}_{N+N_s},
	\end{aligned}
\end{equation}
where $\mathcal{E}_{1/f}$ denotes the sample space of $1/f$ noise, and $\times$ represents the Cartesian product.

By combining Eqs.~\ref{eq-C1}, \ref{eq-C2}, and \ref{eq-C3}, we can express the decay factor $\langle e^{-i\int_0^t\omega^\prime(\tau) d\tau}\rangle_{\varepsilon\in\mathcal{E}}$ as
\begin{widetext}
\begin{equation}\label{eq-C4}
	\begin{aligned}
	\left\langle e^{-i\int_0^t\omega^\prime(\tau) d\tau}\right\rangle_{\varepsilon\in\mathcal{E}}&\equiv\frac{1}{|\mathcal{E}|}\sum_{\varepsilon\in \mathcal{E}}\exp\left(i\int_0^t\omega(\Phi_b+\delta\Phi(\tau,\varepsilon)\right)\\
	&=\exp\left({i\omega_{01}(\Phi_b)t}\right)\bigg(\frac{1}{|\mathcal{E}_{1/f}|}\sum_{\varepsilon_{1/f}}\exp\left(i\int_0^t\frac{d\omega_{01}}{d\Phi_b}\bigg|_{\Phi_b}{\rm Flicker}(\tau,\varepsilon_{1/f})d\tau \right) \\
	&\cdot \frac{1}{N_s|\mathcal{E}^\prime|}\sum_{\varepsilon_1\in \mathcal{E}^\prime}\sum_{\varepsilon_2\in \mathcal{E}^\prime}\cdots \sum_{\varepsilon_{N_s}\in \mathcal{E}^\prime} \prod_{j=1}^{N_s}\exp\left(i\int_0^t\frac{d \omega_{01}}{d\Phi_b}\big|_{\Phi_b}{\rm RTN}_j(\tau,e_j)  d\tau\right)\bigg)\\
	&=\frac{1}{N_s|\mathcal{E}^\prime|}\exp\left({i\omega_{01}(\Phi_b)t}\right)\sum_{\varepsilon_1\in \mathcal{E}^\prime}\sum_{\varepsilon_2\in \mathcal{E}^\prime}\cdots \sum_{\varepsilon_{N+N_s}\in \mathcal{E}^\prime} \prod_{j=1}^{N+N_s}\exp\left(i\int_0^t\frac{d \omega_{01}}{d\Phi_b}\big|_{\Phi_b}{\rm RTN}_j(\tau,e_j)  d\tau\right).
	\end{aligned}
\end{equation}
\end{widetext}

By employing the definition of $\langle \cdot \rangle_{\varepsilon \in \mathcal{E}}$, Eq.~\ref{eq-C4} can be rewritten as
\begin{widetext}
\begin{equation}\label{eq-C5}
	\begin{aligned}
	\left\langle e^{-i\int_0^t\omega^\prime(\tau) d\tau}\right\rangle_{\varepsilon\in\mathcal{E}}&=\exp(i\omega_{01}(\Phi_b)t)\bigg[\left\langle\exp\left(i\int_0^t \frac{d\omega_{01}}{d\Phi_b}\bigg|_{\Phi_b}{\rm Flicker}(\tau,\varepsilon_{1/f})d\tau\right)\right\rangle_{\varepsilon_{1/f}\in \mathcal{E}_{1/f}}\\
	&\cdot
	\prod_{j=1}^{N_s}\left\langle \exp\left(i\int_0^t \frac{d\omega_{01}}{d\Phi_b}\bigg|_{\Phi_b}{\rm RTN_j}(\tau,\varepsilon_j)d\tau\right)\right\rangle_{\varepsilon_j\in \mathcal{E}^\prime} 
	\bigg]\\
	&= \exp(i\omega_{01}(\Phi_b)t) \prod_{j=1}^{N+N_s}\left\langle \exp\left(i\int_0^t \frac{d\omega_{01}}{d\Phi_b}\bigg|_{\Phi_b}{\rm RTN}_j(\tau,\varepsilon_j)d\tau\right)\right\rangle_{\varepsilon_j\in \mathcal{E}^\prime}.
	\end{aligned}
\end{equation}
\end{widetext}

The factor $\exp(i\omega_{01}(\Phi_b)t)$ in Eq.~\ref{eq-C5} represents the dynamical phase in the Schr\"odinger picture. Furthermore, Eq.~\ref{eq-C5} reveals that the decay factor $\langle e^{-i\int_0^t\omega^\prime(\tau) d\tau}\rangle_{\varepsilon\in\mathcal{E}}$ can be decomposed into the product of decay factors contributed by each RTN source.

Based on the analysis of sample functions of RTNs presented in Appendix~\ref{AppendixRamsey}, the term $\langle \exp(i\int_0^t\frac{d\omega_{01}}{d\Phi_b}\big|_{\Phi_b}{\rm RTN}_j(\tau,\varepsilon_j)d\tau)\rangle_{\varepsilon_j\in \mathcal{E}^\prime}$ can be expressed as an expansion, which reads
\begin{widetext}
\begin{equation}\label{eq-C6}
\begin{aligned}
	\left\langle \exp\left(i\int_0^t\frac{d \omega_{01}}{d\Phi_b}\big|_{\Phi_b}{\rm RTN_j}(\tau,\varepsilon_j) d\tau\right) \right\rangle_{\varepsilon \in \mathcal{E}^\prime}
	&\equiv\sum_{n=0}^{+\infty}\sum_{b_0}\int_0^t dt_1\int_{t_1}^t dt_2\cdots\int_{t_{n-1}}^t dt_n  \\&\biggl\{\exp\bigg[i\frac{d\omega_{01}}{d\Phi_b}\big|_{\Phi_b}\int_0^t{\rm RTN}_j(\tau;s_0^\prime,n,t_1,t_2,\cdots,t_n)d\tau\bigg]
	\cdot P(n,b_0,t_1,t_2,\cdots,t_n)\biggr\}.
\end{aligned}
\end{equation}
\end{widetext}
According to the definition in Eq.~\ref{eq-10}, the integration in Eq.~\ref{eq-C6} reads
\begin{widetext}
\begin{equation}\label{eq-C7}
	\int_0^t{\rm RTN}_j(\tau;s_0^\prime,n,t_1,t_2,\cdots,t_n)d\tau=s_0^\prime b_j\left(\sum_{k=1}^n(-1)^{k+1}2t_k+(-1)^n t\right),
\end{equation}
\end{widetext}
where $b_j$ denotes the amplitude of the $j$-th RTN for $1 \leq j \leq N+N_s$, the parameters $s^\prime$, $n$, and $\{t_j\}$ are defined in Appendix~\ref{AppendixRamsey}. Furthermore, the probability distribution function in Eq.~\ref{eq-C6} is given by
\begin{widetext}
\begin{equation}\label{eq-C8}
	\begin{aligned}
	P(n,s_0^\prime,t_1,t_2,\cdots,t_n)&=P(s_0^\prime)P(t_1,t_2,\cdots,t_n|n)P(n)\\
	&=P(s_0^\prime)\cdot\frac{n!}{t^n}\cdot\frac{(\lambda_j t)^n e^{-\lambda_j t}}{n!}\\
	&=P(s_0^\prime)e^{-\lambda_j t}{\lambda_j}^n,
	\end{aligned}
\end{equation} 
\end{widetext}
where $P(s_0)$ follows a discrete distribution such that $P(s_0^\prime = +1) = P(s_0^\prime = -1) = 1/2$, $\lambda_j$ represents the switch frequency of the RTN.

By substituting Eqs.~\ref{eq-C7} and \ref{eq-C8} into Eq.~\ref{eq-C6}, we derive a formal expression for the decay factor caused by a single RTN, which is given by
\begin{widetext}
\begin{equation}\label{eq-C9}
\begin{aligned}
	&\left\langle \exp\left(i\int_0^t\frac{d \omega_{01}}{d\Phi_b}\big|_{\Phi_b}{\rm RTN}_j(\tau,\varepsilon_j) d\tau\right) \right\rangle_{\varepsilon_j\in \mathcal{E}^\prime}\\
	&=\frac{1}{2}e^{-\lambda_j t}\sum_{n=0}^{+\infty}\int_0^t dt_1\int_{t_1}^t dt_2\cdots\int_{t_{n-1}}^t dt_n \biggl\{\exp\bigg[i\frac{d\omega_{01}}{d\Phi_b}\big|_{\Phi_b}b_j \bigg(\sum_{k=1}^n(-1)^{k+1}2t_k+(-1)^n t\bigg)\bigg]\\
	&+\exp\bigg[-i\frac{d\omega_{01}}{d\Phi_b}\big|_{\Phi_b} b_j\bigg(\sum_{k=1}^n(-1)^{k+1}2t_k+(-1)^n t\bigg)\bigg]\biggr\}{\lambda_j}^n.
\end{aligned}
\end{equation}
\end{widetext}

According to Eq.~\ref{eq-C5}, the total decay factor $\langle e^{-i\int_0^t \omega^\prime (\tau)d\tau}\rangle_{\varepsilon\in \mathcal{E}}$ is simply the product of all RTNs' contribution. Therefore, if the RTNs can be characterized on a case-by-case basis, Eq.~\ref{eq-C9} would serve as a powerful tool for evaluating decoherence phenomena induced by either $1/f$ noise or a few strong RTNs.

\bibliographystyle{apsrev4-1} 
\bibliography{Bibfile}

\end{document}